# Strain multiplexed metasurface holograms on a stretchable substrate

*Stephanie C. Malek[†], Ho-Seok Ee[†], and Ritesh Agarwal[*]*

Department of Materials Science and Engineering, University of Pennsylvania, Philadelphia, PA 19104, USA.

ABSTRACT

We demonstrate reconfigurable phase-only computer-generated metasurface holograms with one, two, or three image planes operating in the visible regime on a stretchable polydimethylsiloxane substrate. Stretching the substrate enlarges the hologram image and changes the location of the image plane. Upon stretching, these devices can switch the displayed holographic image between multiple distinct images. This work opens up the possibilities for stretchable metasurface holograms as flat devices for dynamically reconfigurable optical communication and display. It also confirms that metasurfaces on stretchable substrates can serve as platform for a variety of reconfigurable optical devices.

Metasurfaces are flat nanostructured surfaces with subwavelength thicknesses, and many are designed so as to shape an optical wavefront on subwavelength length scales in order to mimic or improve upon the functionalities of bulk refractive optical components.[1] While conventional optical components rely on phase accumulation through propagation, metasurfaces depend on the spatial variation of phase discontinuities induced by an ensemble of carefully designed nanoscale resonators that each induce a deliberate phase discontinuity to the incident optical wavefront. The specific local phase shift (0-2π) caused by each resonator can be engineered based on its shape,[2–4] dimensions,[5,6] or geometrical orientation angle.[7,8] Many metasurface devices have been developed including demonstrations of a generalization of Snell's law,[2,3,7,9] flat lenses,[6,8,10–13] quarter wave plates,[4] and vortex beam generation.[7,14] Metasurface holograms[15–17] are of particular interest for communication and information storage. Advancements in metasurface holograms include multi-color holograms[18–20] and three dimensional holograms.[21,22] Several metasurface holograms have also been demonstrated such that the holographic image can be changed by switching the polarization of incident light.[5,22–24] These devices require adjustment of bulk optical components to change the polarization of the incident light and can only switch between two holographic images. Truly reconfiguring the functionality of a metasurface in a dynamic and self-contained manner requires electrically modifying the material properties of individual nanostructures[25] or changing the relative positions of nanostructures with applied strain.[26,27] Here, we introduce metasurface holograms on stretchable substrates that can be continuously reconfigured by isotropic stretching. These metasurfaces, upon stretching, can multiplex between two or more holographic images.

In our previous work on tunable metasurfaces on stretchable substrates[27], we established the ability to continuously tune a wavefront by stretching a metasurface. Most notably, we showed a metasurface zoom lens on a polydimethylsiloxane (PDMS) substrate whose focal length can be changed up to ~1.7x when stretched by ~30%. In this work, we follow the same principle: isotropic stretching of the metasurface alters the position dependent phase discontinuity and reconfigures the resulting optical wavefront. The electric field profile, $E_0$, of transmitted light on an unstretched metasurface can be written as $E_0(x, y) = Ae^{ik\Phi(x,y)}$. When the metasurface is stretched isotropically by a factor of $s$, the electric field profile changes to $E_0'$, which is related to the original profile as $E_0'(sx, sy) = E_0(x, y)$. To understand rigorously how a wavefront changes with stretching, we employ a Fresnel transformation to consider the electric field distribution of light shaped by a metasurface.[28] From Huygens-Fresnel Principle, the electric field $E(x, y, z)$ at a distance $z$ from a metasurface can be expressed as

$$E(x, y, z) = \iint E_0(x_0, y_0)(\tfrac{-ik}{2\pi z})e^{ikz}e^{ik[(x-x_0)^2+(y-y_0)^2]/2z}dx_0 dy_0 \qquad (1)$$

We can then derive that, for a given stretch ratio $s$, the electric field is modified by stretching such that

$$E'(sx, sy, s^2 z) = E(x, y, z)e^{ik(s^2-1)z} \qquad (2)$$

If a metasurface is stretched by a factor of $s$, the electric field—at any plane that is a factor of $s^2$ farther away from the metasurface than its unstretched position—will be expanded by a factor of $s$ in the $x$ and $y$ directions compared to its corresponding unstretched field distribution. This relation applies broadly to stretchable metasurfaces, of which a lens is a simple example. As the metasurface lens is stretched, its focal length changes such that the new focal length $f'$ can be defined as $f' = s^2 f$ where $f$ is the original focal length.[27] Accordingly, if a metasurface hologram is stretched by a factor of *s*, the hologram image is also stretched by a factor of *s*, and the hologram image plane moves in the *z* direction away from its unstretched *z* position as $s^2$ (Figure 1a).

In order to verify this behavior experimentally, we designed, fabricated, and evaluated metasurface holograms on stretchable polydimethylsiloxane (PDMS) substrates. We used a computer generated hologram (CGH) technique based on Fresnel's ping-pong algorithm to determine the phase distribution needed to reconstruct a single holographic image[29] or a multi-plane hologram.[30] In our devices, we generated this phase distribution with the position dependent orientation angle of anisotropic gold nanorods following the notion of Pancharatnam-Berry phase.[31,32] Each nanorod acts as a quarter waveplate to convert the circular polarization of incident light, A metasurface comprised of anisotropic resonators whose orientation angle varies with position as $\varphi(x, y)$ will also induce a position dependent phase discontinuity $\Phi(x, y)$ that depends only on the orientation angle of the resonator such that $\Phi(x, y) = \pm 2\varphi(x, y)$ where + (-) represents right (left) circularly polarized incident light. In our metasurfaces, gold nanorods with dimensions of *w*=90 nm, *l*=220 nm, *h*=70 nm were arranged in a hexagonal lattice with a lattice constant *a* of 340 nm or 360 nm for maximal cross-polarization transmission efficiency.[27] To examine device behavior, we fabricated and measured these metasurface holograms following a similar procedure as the one outlined in our previous work on stretchable metasurface zoom lenses[27] (see Supporting Information for Methods). In short, we fabricated metasurfaces on a silicon handler wafer (SEM image: Figure 1b) before transferring them to stretchable PDMS substrates (Optical images: Figure1c and d). As a function of stretch ratio *s*, we experimentally measured the distance between the metasurface and the hologram image plane.

We first designed a metasurface hologram with an image, "PENN" (Figure 2a), located 150 μm away from the metasurface using the CGH technique.[29] Figure 2b is a numerically reconstructed hologram image from the obtained phase distribution and agrees well with the original image. We fabricated this device on a PDMS substrate and evaluated its behavior with stretching. As the metasurface is stretched, the hologram undergoes two changes. First, the size of the hologram image increases in proportion to the stretch ratio $s$. This is revealed by a visual comparison of Figure 2c and d insofar as the hologram resulting from the stretched metasurface ($s$=1.24) in Figure 2d is 1.24 times larger than for the unstretched metasurface (Figure 2c). Second, the hologram image plane moves away from the metasurface as a function of $s^2$. For example, when this metasurface is stretched such that $s$=1.24, the hologram image plane is predicted to be 230.6 μm (measured: 232 μm) away from the metasurface compared to 150 μm (measured: 152 μm) for $s$=1. Figure 2e shows reasonable agreement between measured and calculated values for both the stretch ratio of the hologram image and the distance between the metasurface and the hologram image plane as a function of $s$ the stretch ratio of the metasurface. It is therefore clear that the behavior of these metasurface holograms on stretchable substrates can be modified predictably with stretching.

We leverage the fact that the image plane position changes when the device is stretched to develop reconfigurable metasurface holograms where the hologram image observed at a fixed plane changes upon stretching. We demonstrate this concept with a stretchable metasurface hologram metasurface that can switch between two very different images. Using computational techniques for generating multi-plane holograms,[30] we designed a two-plane metasurface hologram that has, in its unstretched form, an image plane with the word "ONE" (numerical reconstruction: Figure 3a) located 130 μm from the metasurface and another image plane with the word "TWO" (Figure 3b) located 200 μm from the metasurface. We fabricated this device on a PDMS substrate and observed the hologram image at a plane 200 μm from the metasurface as a function of isotropic stretching. As the metasurface (located at $z$=0) is stretched, both image planes move away from the metasurface in proportion to $s^2$. As shown in Figure 3c, when the device is stretched sufficiently ($s$=1.24), the image plane with the word "TWO" moves to $z$=200 μm. In this way, the hologram image observed at $z$=200 μm switches from "ONE" to "TWO" with stretching. Figures 3d and e show measurements taken at fixed location of $z$=200 μm for $s$=1 (Figure 3d) and $s$=1.24 (Figure 3e) and clearly demonstrate that the hologram image changes from "ONE" to "TWO" with isotropic applied strain. As further confirmation of this concept of switchable metasurfaces, we designed, fabricated, and measured a metasurface hologram on PDMS comprised of anisotropic silicon nanorods (instead of gold nanorods). In its unstretched form, this device has the image plane with an image of a smiling face at $z$=60 μm and an image of a frowning face at $z$=100 μm. Upon stretching such that $s$=1.3, the holographic image at $z$=100 μm switches from a frowning face to a smiling face (Supporting Information Figure S1). As such, we have experimentally realized metasurface holograms that are switchable with isotropic applied strain.

To show that switchable behavior is not limited to holograms with only two image planes we develop a stretchable metasurface (with gold nanorods) that can switch between three simple but distinct images. The performance of such a device can be degraded by noise that results from interference between image planes, and therefore care was taken to place image planes sufficiently far apart and adjust appropriately the relative intensity of the three holographic images. We designed a metasurface hologram with three planes such that, for an unstretched device, there are image planes at $z$=340 μm, $z$=270 μm, and $z$=200 μm with images of a pentagon (numerical reconstruction: Figure 4a), a square (Figure 4b), and a triangle (Figure 4c) respectively. When the metasurface is stretched, the hologram image at $z$=340 μm changes from a pentagon ($s$=1) to a square ($s$=1.12) to a triangle ($s$=1.30), as shown experimentally in Figures 4d, e, and f. The operating principle for this metasurface is the same as the two-plane switchable hologram in that each image plane moves away from the metasurface as $s^2$ as the metasurface is stretched. We measure the distance between the metasurface and each hologram image (Figure 4g) and confirm good agreement between predicted and measured values of image plane location as a function of stretch ratio for each of the image planes. A similar metasurface with image planes at $z$=220 μm (triangle), $z$=290 μm (square), and $z$=360 μm (pentagon) was fabricated, and the image plane at $z$=360 μm at 35 different stretch ratios was recorded to show the transition between different image with stretching. As expected, the hologram image at $z$=360 μm changes continuously as it is stretched (see Supporting Information Movie S1). From these two devices, we demonstrate stretchable metasurface holograms that, on a fixed observation plane, can multiplex between three entirely different images with stretching.

In summary, we have demonstrated metasurface holograms on stretchable substrates where the hologram image can be multiplexed by stretching. A stretchable metasurface hologram capable of multiplexing through many different holographic images could serve as a compact means of conveying a significant amount of information and might also enable holograms that can be animated by stretching. Development of high efficiency and polarization insensitive stretchable metasurface holograms with dielectric (Si, $TiO_2$), as opposed to plasmonic resonators, could be a substantial advance. Additionally, multicolor stretchable metasurface holograms and stretchable non-linear metaurface holograms may enable practical applications of this device. Stretchable metasurface holograms may prove useful for applications such as virtual reality, flat displays, and optical communication.

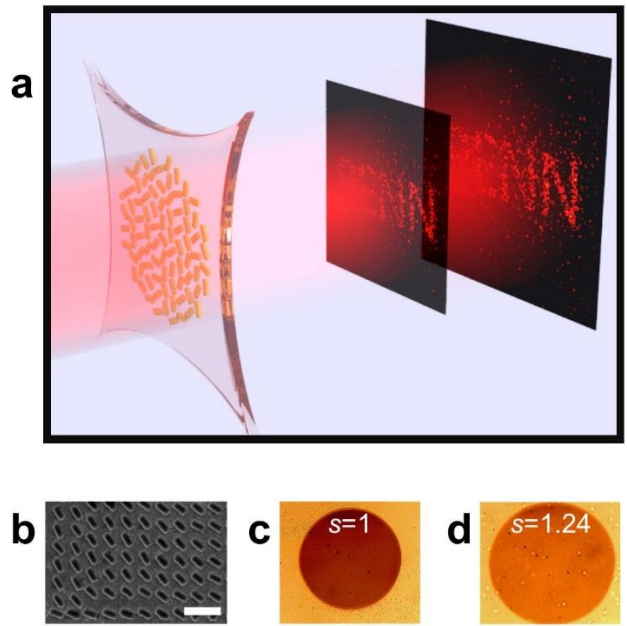

**Figure 1**. Overview of stretchable metasurface hologram device. (a) Schematic of single plane stretchable hologram. (b) Scanning Electron Microscope (SEM) image of a device on silicon before transfer to a PDMS substrate. Scale bar is 1 μm (c-d) Optical bright field images of a device when (c) unstretched and (d) stretched 124%.

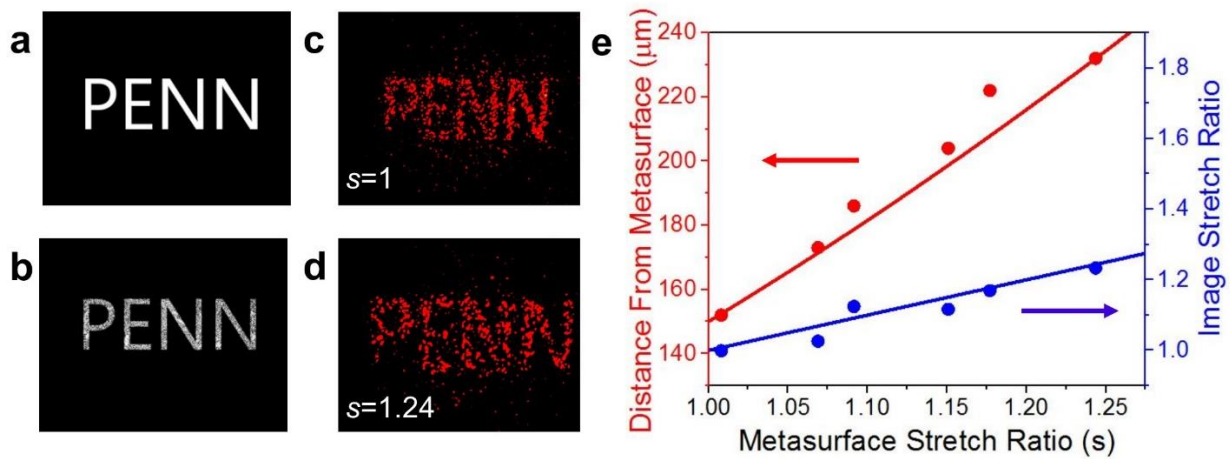

**Figure 2**. Design and performance of a single plane metasurface hologram on a stretchable substrate. (a) Original image used to generate phase distribution for metasurface hologram. (b) Computer reconstruction of hologram image. (c-d) Experimentally obtained hologram images for an (c) unstretched and (d) stretched device (exposure time is 1.67x longer than (c)). (e) Plot of calculated (solid lines) and measured (dots) image stretch ratio (blue) and image plane location (red) as a function of device stretch ratio.

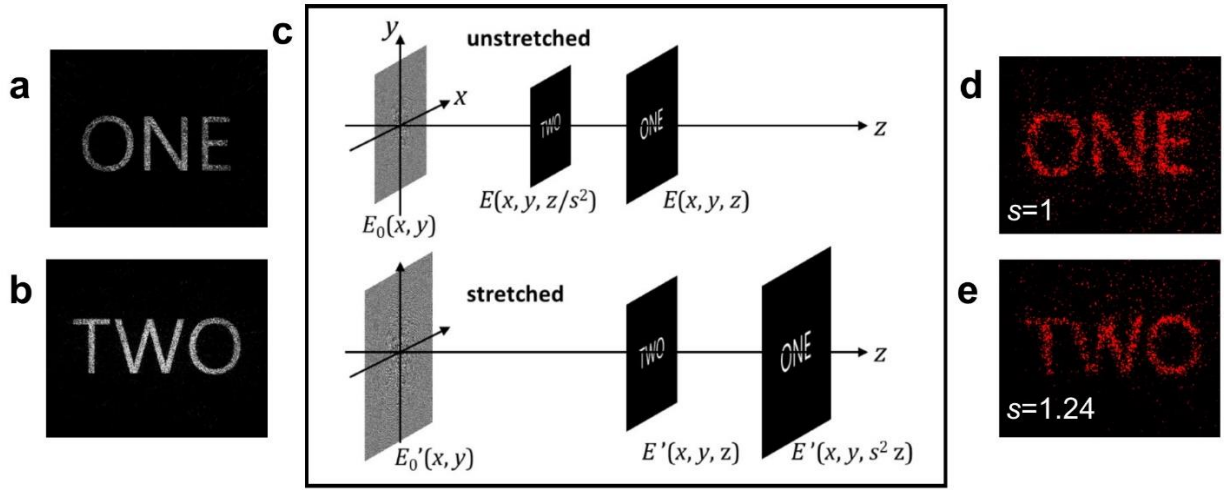

**Figure 3**. Design and performance of a two plane metasurface hologram on a stretchable substrate. (a-b) Computer reconstructions of computer generated hologram image planes (c) Schematic of two-plane hologram and mechanism for hologram image switching by stretching. (d-e) Experimentally measured hologram images at an image plane 200 μm from the metasurface for a devices that is (d) not stretched and (e) stretched 124%.

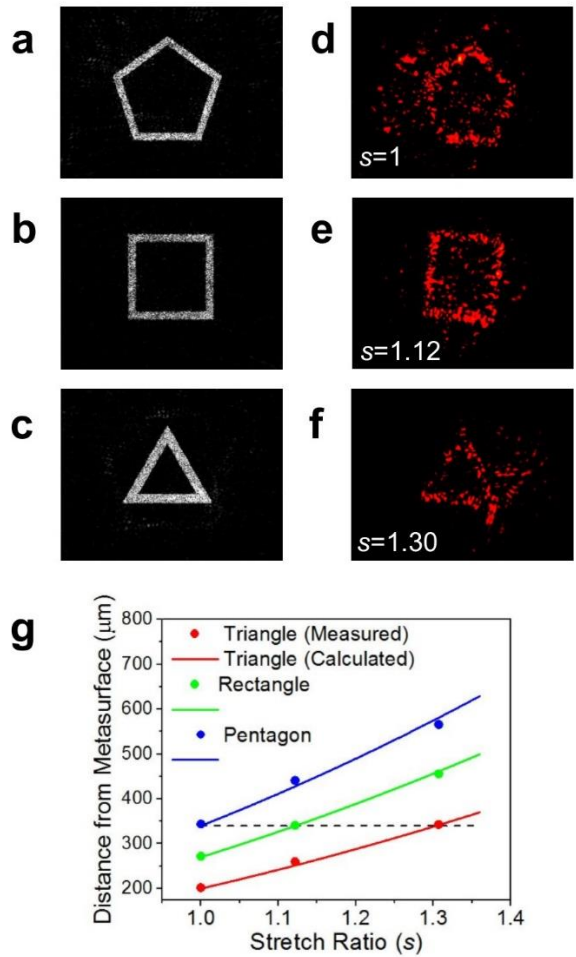

**Figure 4**. Design and performance of a three plane metasurface hologram on a stretchable substrate. (a-c) Computer reconstructions of computer generated hologram image planes. (d-f) Experimentally captured hologram images at a distance of 340 μm from the metasurface when the device is (d) unstretched, (e) stretched 1.12%, and (f) stretched 130%. (g) Plot of calculated (solid lines) and measured (dots) image plane location as a function of stretch ratio.

## ASSOCIATED CONTENT

**Supporting Information**. Methods for device fabrication and optical measurement. Performance of a switchable silicon metsurface hologram device on PDMS. A movie showing operation of stretchable, switchable metasurface hologram. Those materials are available free of charge via the Internet at http://pubs.acs.org.

## AUTHOR INFORMATION

**Corresponding Author**

*riteshag@seas.upenn.edu

**Author Contributions**

†These authors contributed equally.

**Notes**

The authors declare no competing financial interest.

## ACKNOWLEDGMENT

This work was supported by the U.S. Army Research Office MURI grant (Grant # W911NF-11-1-0024). S.C.M acknowledges the NSF Graduate Research Fellowship Program for its support.